# An investigation of the 'standard equation' for the energy hoarded in an Elastic Material during elongation and a bid to crack the anomalies by including the parameters neglected for deriving the 'standard equation'

Aasis Vinayak.P.G.

Varadakshina, Perayam, Mulavana.P.O.,

Kollam, Kerala - 691503, India

Abstract

Paper examines the validity and soundness of the standard equation derived to find the amount of energy stored inside an elastic material when it is stretched. The paper also tries to incorporate the parameters that where neglected while deriving the so-called 'standard equation' and thus by trying to solve the anomalies associated with it. The paper further suggests a new equation which can answer all the questions.

## 1. Introduction

Let us begin the paper with a thought experiment.

Consider the wire of length *l* and area of cross section *A* fixed vertically at one end and at the other end we hang a substance of mass m. Due to gravitational force it will exert a force of *mg* (if g is the acceleration due to gravity at that point) downward. This produces some extension in the wire. Consequently the weight will also come down by an equal distance to that of the change in length. Here we are making an assumption that the elongation *x* (see the figure) is so small so that the change in g, $\Delta g \to 0$ and thus g will remain constant (this is done for simplicity even other wise also we can do the similar mathematics using the relation for its declining rate and the value we get will be very small as the elongation x is small – even when one resorts to get shelter here he could account for the half of energy but will get only a meager portion).

The energy stored in the elastic material, as per the classical equation [1], is

---

Electronic mail: aasisvinayak@gmail.com

*E = ½ x Stretching force x extension* (1)

And here it is

$$E = \frac{1}{2} \times mg \times x \qquad (2)$$

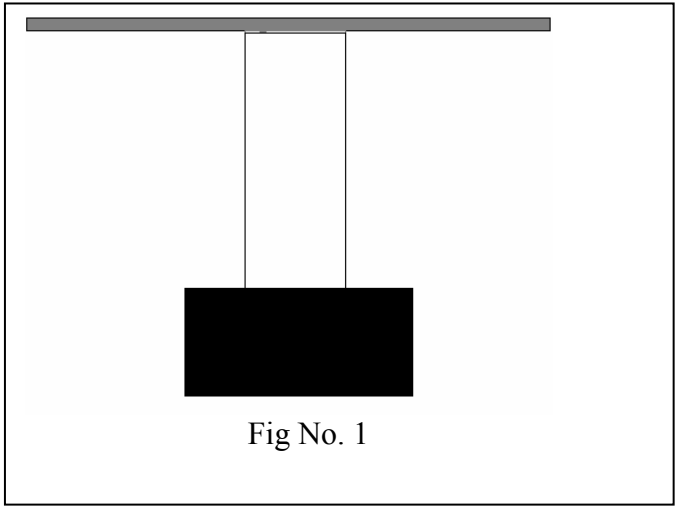

Fig No. 1

The decrease in gravitational potential energy is given by

$$E = mg \times x \qquad (3)$$

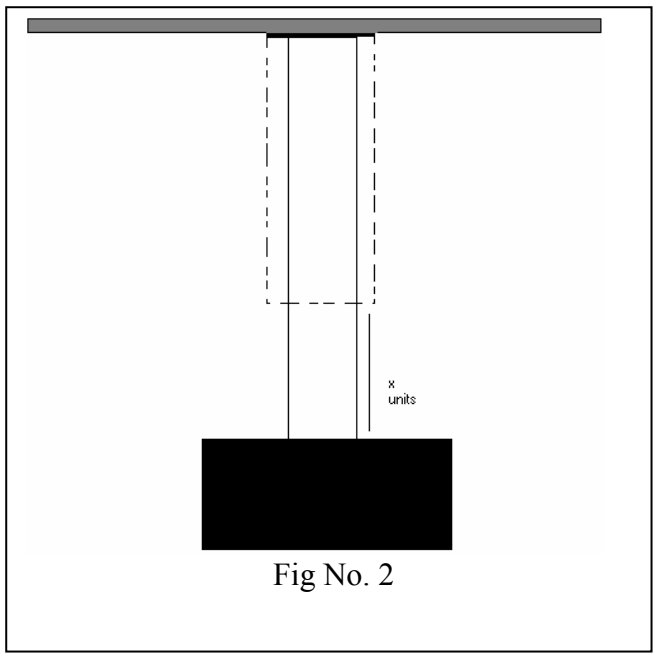

Fig No. 2

The equation (1) and (2) shows an unbalance and thus the classical standard equation fails to include the half of the energy. Thus it cannot answer what happened to the remaining half the energy.

## 2. Analysis

In the classical derivation the change in area has been neglected. Let us include that also in our purview:

*Case No.1*

In the case consider the classical way in which we derived the equation [1] where we have neglected the change in area (also considered an ideal situation (which is not relevant to mention here) where there is no dissipation of energy and the Young's Modulus $Y$ will remain as a constant during extension as well during compression – this is followed in the second case also) but considered only the change in length and thus we got the equation for the energy stored as (1)

*Case No.2*

Consider a similar case with area, i.e., here the change in length is not taken into account as it has been already been referred and found (or can be simply stated as neglected – when combined together with the *case no. 1* no parameter will be neglected). [1]

Here as we are considering the ideal case, the total volume is remaining constant, the same condition we applied to find the maximum value of Poisson's ratio ($\sigma$) of a material under ideal conditions. So here the value of $\sigma$ will always be that in the ideal case, i.e., equal to 0.5 [2, 3]

Let us represent the force acting downward as
$$F = mg \qquad (4)$$
And that acting in the lateral direction as F'
It is indispensable to find an expression for F' in terms of F. For that reflect on this case -
We have the standard relation for the longitudinal elongation [4, 5] as
$$Y = \frac{F}{A_1} \times \frac{L}{l} \qquad (5)$$
, where $Y$ is the Young's modulus, $F$ is the force acting longitudinally, $A_1$ is the area of cross section of the end, $L$ is the initial length and $l$ is the change in length.
And for the lateral compression (as it is not acting normally is all directions) we have [1]

$$Y = \frac{F'}{A_2} \times \frac{R}{r} \qquad (6)$$
, where $F'$ is the force acting laterally, $A_2$ is the lateral area of the wire, $R$ is the initial radius and $r$ is change in radius.

We have the values
$$A_1 = \pi R^2 \tag{7}$$
$$A_2 = 2\pi RL \tag{8}$$

Equating (5) and (6) we get

$$\frac{F}{A_1} \times \frac{L}{l} = \frac{F'}{A_2} \times \frac{R}{r} \tag{9}$$

Putting the values (7) and (8)

$$\frac{F}{\pi R^2} \times \frac{L}{l} = \frac{F'}{2\pi RL} \times \frac{R}{r} \tag{10}$$

$$F' = \frac{2 \times F \times L^2 \times r}{l \times R^2} = \frac{2 \times F \times L \times \frac{r}{R}}{\frac{l}{L} \times R} \tag{11}$$

, which can be re-written as

$$F' = 2 \times F \times \frac{L}{R} \times \frac{\frac{r}{R}}{\frac{l}{L}} \tag{12}$$

And we know for ideal case like this

$$\frac{\frac{r}{R}}{\frac{l}{L}} = \sigma = \frac{1}{2} \tag{13}$$

, which implies that

$$F' = F \times \frac{L}{R} \tag{14}$$

Now consider the fig No 3 (which shows the aspects of lateral contraction) when there is-

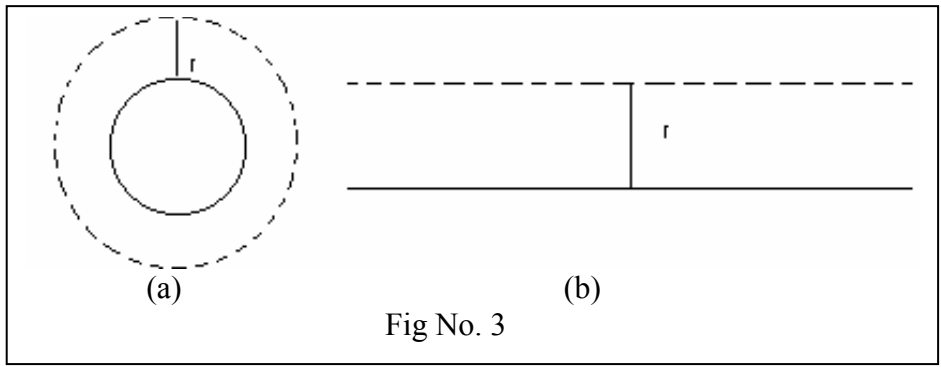

(a)          (b)

Fig No. 3

-a transformation from (a) to (b) there wouldn't be any change in conditions [6]. From (b) we can say that the work done is

$$W = F' \times r \qquad (15)$$

From (14) is equal to

$$W = \left(F \times \frac{L}{R}\right) \times r \qquad (16)$$

From (13) we can say

$$\frac{L}{R} \times r = \frac{1}{2} \times l \qquad (17)$$

Putting this in (16) we get

$$W = \frac{1}{2} \times F \times l \qquad (18)$$

## 3. Conclusion

The equation for the energy stored in an elastic material during elongation is given by

$$W = F \times l \qquad (19)$$

Thus in total, when both the cases (length as well as area) are considered we will get the total energy stored (19) and there by we could answer any question concerning the law of conservation of energy with respect to this system.